\documentclass[aps,prd,twoside,twocolumn,superscriptaddress,floatfix,nofootinbib,showpacs]{revtex4-2}
\usepackage{amsmath}

\usepackage{booktabs,tabularx}
\newcolumntype{Y}{>{\centering\arraybackslash}X}

\usepackage{bm}
\usepackage{booktabs}
\usepackage{xcolor}
\usepackage{hyperref}
\usepackage{url}
\usepackage{graphicx}
\usepackage[large]{subfigure}
\usepackage{amssymb}
\usepackage[amssymb]{SIunits}

\usepackage{natbib}
\usepackage{multirow}
\usepackage{makecell}

\usepackage{xcolor}
\usepackage{soul}

\usepackage{cleveref}

\usepackage{simpler-wick}

\usepackage{verbatim}

\usepackage{color}

\usepackage{bm}

\usepackage{hyperref}
\usepackage{url}
\usepackage{graphicx}
\usepackage[large]{subfigure}
\usepackage{amssymb}
\usepackage[amssymb]{SIunits}

\usepackage{natbib}
\usepackage{multirow}
\usepackage{amsmath}

\usepackage{cleveref}
\usepackage{simpler-wick}

\newcommand{\picocmb}{{\texttt{PICO}}}
\newcommand{\litebird}{{\texttt{LiteBIRD}}}

\newcommand{\apjs}{Astrophys. J. Suppl. Ser.}
\newcommand{\aap}{Astron. Astrophys.}

\newcommand{\jcap}{J. Cosmol. Astropart. Phys.}
\newcommand{\mnras}{Mon. Not. R. Astron. Soc.}

\renewcommand\prd{{\rm Phys. Rev. D}}
\renewcommand\prl{{\rm Phys. Rev. Lett.}}

\newcommand{\healpix}[1]{\texttt{HEALPix} #1}
\newcommand{\namaster}[1]{\texttt{NaMaster} #1}

\begin{document}
\title{Inference of $B$-mode polarization in the presence of non-Gaussian foregrounds}

\author{Sen Li}
\affiliation{Department of Astronomy,
University of Science and Technology of China, Hefei 230026, China}
\affiliation{School of Astronomy and Space Science, University of Science and Technology of China, Hefei 230026, China}

\author{Chang Feng}
\altaffiliation{Corresponding author: changfeng@ustc.edu.cn}
\affiliation{Department of Astronomy,
University of Science and Technology of China, Hefei 230026, China}
\affiliation{School of Astronomy and Space Science, University of Science and Technology of China, Hefei 230026, China}

\author{Filipe B. Abdalla}
\affiliation{Department of Astronomy,
University of Science and Technology of China, Hefei 230026, China}
\affiliation{School of Astronomy and Space Science, University of Science and Technology of China, Hefei 230026, China}

\begin{abstract}
The inflationary $B$-mode signals encode invaluable information about the origin of our Universe and searching for potential signatures of primordial gravitational waves (PGWs) is one of the major science goals for future precision observations of cosmic microwave background (CMB) polarization. However, dominant $B$-mode signals of both Galactic foreground contamination and gravitational lensing effects prevent direct measurements of the PGW $B$-mode signals. There are existing proposals which can effectively eliminate these two contaminants but issues remain for future high-sensitivity and multifrequency CMB polarization observations, such as spatially-varying spectral energy distribution (SED) of polarized foreground and cosmological $B$-mode signals due to primordial magnetic fields (PMFs). In this work, we investigate inference of PGW $B$-mode signals in the presence of both complexities. We employ a constrained moment internal linear combination (cMILC) method to remove polarization signals arising from spatially varying SEDs. Also, we employ a power-spectrum-based approach to extracting both the Galactic and cosmological $B$-mode components. Two methods have been validated by mock data and different consistency tests have been performed. We apply these two methods to end-to-end simulations for future high-sensitivity and multifrequency polarization observations and investigate the detectability of different $B$-mode signals in the presence of non-Gaussian polarized foregrounds under different scenarios. This study will be important for new physics studies with $B$-mode signatures.
\end{abstract}

\maketitle

\section{Introduction}
\label{intro}

Future observations of cosmic microwave background (CMB) polarization will achieve unprecedented precision to map both primary and secondary fluctuations of electric-like $E$-mode and curl-like $B$-mode polarization. The primary $B$-mode signals could be generated by physical processes in the early Universe such as inflationary mechanisms~\cite{1997PhRvL..78.2058K, 1997PhRvL..78.2054S}, parity-violating interactions~\cite{1998PhRvL..81.3067C,1999PhRvL..83.1506L} and primordial magnetic fields (PMFs)~\cite{2004PhRvD..70d3011L, 2014PhRvL.112s1303B}, whereas the secondary $B$ modes can originate from gravitational lensing effects~\cite{1998PhRvD..58b3003Z}, anisotropic cosmic rotation effects~\cite{2009PhRvL.102k1302K} and patchy reionization~\cite{2007PhRvD..76d3002D,2009PhRvD..79j7302D}. 

Measuring primordial gravitational waves (PGWs) is one of the major goals for current and future CMB experiments~\cite{so,cmbs4}. However, polarized Galactic foreground and CMB lensing can significantly contaminate the PGW $B$-mode signals at both the large and small angular scales~\cite{so_delensing, cmbs4_delensing}. 

Moreover, tensor perturbations of the PMFs can generate $B$-mode signals with a similar shape as the PGW at large angular scales. Thus, the PMF-induced $B$ modes may further contaminate the PGW ones, indicating that there would be a strong degeneracy between the tensor-to-scalar ratio and the PMF amplitude even in the absence of polarized Galactic foreground~\cite{degeneracy, degeneracy2}. The degeneracy can be reduced or even broken by combing the $B$-mode measurements at both large and small angular scales. Meanwhile, the PMF can induce Faraday rotation of the CMB polarization so the anisotropic rotation angle measurements would further break the degeneracy.  

However, the PMF signatures may be too faint to be detected if its strength is below $\sim$0.1 nano-Gauss as implied by recent analyses~\cite{levon1, levon2} and the $B$-mode signals due to unresolved PMF may consequently bias that of the PGWs. Different from the potential $B$-mode contamination arising from cosmological origins such as the PMF, the polarized Galactic foreground is another potential challenge. 

The $B$-modes of polarized dust and synchrotron emission have been detected by BICEP/Keck experiment at 95 and 150 GHz, respectively~\cite{bkbmode}. Also, the recent $B$-mode measurement from a wide patch by POLARBEAR experiment revealed evidence for Galactic dust polarization at 150 GHz~\cite{pbbmode}. Moreover, both the measured $TB$ and $EB$ cross-power spectra by Planck experiment at high frequencies are nonzero, indicating a complex nature of the polarized foreground~\cite{pktb}. It is worth noting that a theory of a misalignment angle between directions of Galactic filamentary dust and magnetic fields was recently proposed to explain these nonzero $TB$ and $EB$ cross-power spectra, and the misalignment angle can be determined by the measured $TB$ and $TE$ cross-power spectra~\cite{theory_tbeb}. In addition, the nonzero $EB$ cross-power spectrum of the polarized foreground could also contaminate a cosmological $EB$ cross-power spectrum due to a polarization rotation by axion-like particles~\cite{MK2020} and should be separated from the raw $EB$ cross-power spectrum by adopting the misalignment angle model~\cite{ebmodel}. 

A nonvanishing cross-power spectrum $EB$ is a very interesting probe to parity-violating physics which predicts that the polarization angle could be rotated and cause a cosmological rotation effect~\cite{1998PhRvL..81.3067C,1999PhRvL..83.1506L}. Especially, the recent measurement using the Planck CMB data seems to detect a nonzero rotation angle at $0.3^{\circ}$~\cite{MK2020}. The key to extracting such an angle is to include the foreground polarization signals which have different power-spectrum shapes from the CMB so the degeneracy between the cosmological and instrumental rotation signals can be broken. However, if the non-Gaussianity of the polarized foreground exists and is unaccounted for, it could affect the $EB$ correlations. 

The Planck observations have found that the foreground amplitudes and spectral indices are spatially varying~\cite{pkng}. The measured correlation coefficients among high frequency channels would be inconsistent with theoretical predictions if the spatially varying spectral indices are not taken into account. 

The spatially varying SEDs can induce extra foreground fluctuations in the CMB polarization signals and require a delicate foreground removal procedure. The internal linear combination (ILC) method is a standard approach to extracting the CMB signals by identifying its black body spectrum~\cite{ilc1, ilc2} but can not fully suppress the foreground. The constrained ILC (cILC) is more advanced since it imposes an additional constraint on the ILC weighting for nulling out the foreground component~\cite{cilc}. Both the ILC and cILC methods assume that the SEDs are isotropic.
If the spatial fluctuations of the foreground SEDs are sub-dominant, we can perturbatively expand the SEDs around the isotropic ones so the SED spatial components can be considered as new foreground species whose SEDs are derivatives of the isotropic SEDs. While several component separation techniques have been proposed for CMB polarization analysis, including template subtraction, parametric modeling (e.g., Commander~\cite{commander}), and non-parametric approaches such as generalized needlet ILC (GNILC)~\cite{gnilc}, the cMILC method offers a distinctive advantage by analytically incorporating spatial variations in the SEDs of polarized foregrounds. Unlike GNILC, which relies on signal-to-noise weighting in needlet space and can struggle with structured non-Gaussian foregrounds, cMILC directly imposes moment constraints on SED derivatives, allowing for explicit nulling of spatial SED fluctuations. Moreover, cMILC preserves the semi-blind nature of the ILC family while extending its flexibility to account for physically motivated foreground complexity, all without requiring strong priors or external templates. This makes cMILC particularly well-suited for next-generation polarization experiments, where precision control of foreground residuals across many frequency channels is essential.

Different from the cMILC which is an ILC variant in map space, an alternative approach which works in power-spectrum space has been proposed as well~\cite{psbased}. The foreground has to be removed before the cosmological signals can be detected if the map-space method is adopted. The power-spectrum-space approach does not require the sequential steps and can directly extract all the key information for the multifrequency power spectra formed at different frequencies. Although the SED spatial components are not detected from BICEP/Keck $B$-mode power spectra because of the realistic sensitivity and sky coverage for the polarization measurements~\cite{psbased}, the higher-order power spectra due to the spatial SED components may not be negligible for future high-sensitivity polarization measurements and would cause significant biases on the PGW inference if unaccounted for.

Both the map-space-based and powers-spectrum-space-based methods have been applied to different datasets in the literature. The cMILC method~\cite{taylormoment} has been investigated among other methods for the tensor-to-scale ratio forecast of the Simons Observatory~\cite{so23}. The moment-expansion method with multifrequency power spectra has been applied to the Planck datasets~\cite{mangilli19}, the BICEP/Keck datasets~\cite{psbased}, and the simulated datasets for future space-borne CMB observations~\cite{momentexp_litebird, momentexp2}. In addition, the cMILC method is further upgraded into the optimized cMILC (ocMILC)~\cite{ocmilc} in light of the partial ILC method introduced in~\cite{pilc}.

The future CMB polarization will have ultra-high sensitivity and cover a broad range of frequencies and significant polarized foreground will be present at both low and high frequencies. Especially, the non-Gaussian fluctuations due to the spatially-varying SEDs may induce nonnegligible higher-older correlations which can manifest similar descending features degenerate with the large-angular-scale cosmological signals, further complicating the known degeneracy between the PGW and PMF signals due to the similar power-spectrum shapes at large angular scales~\cite{degeneracy, degeneracy2}.
These degeneracies among the polarized foreground, PMF and PGW would make the detection of the cosmological signals more challenging.

In this work, we for the first time investigate whether the strong degeneracy between PMF and PGW can still be broken when there is a bright and complex foreground contamination with a non-negligible non-Gaussianity. The motivation of this study is to focus on the application of established component separation techniques and methods in the joint inference of PGW and PMF $B$-mode signals in the presence of non-Gaussian foregrounds. Actually the latter has been studied in the literature~\cite{degeneracy,degeneracy2} where a strong degeneracy between PMF and PGW has been found without considering any foreground contamination issues. 

The structure of this work is as follows. We describe the microwave sky model in Sec. II, and discuss different foreground methods and validation results in Sec. III. We conclude in Sec. IV.

\section{CMB Sky model}
\label{skysec}
The CMB data structure can be described as 
\begin{eqnarray}
    X^{\rm obs}_{\nu}(\bf n)&=&a_{\nu}X^{\rm CMB}({\bf n})+f_{\nu}{\tilde A}({\bf n})+N_{\nu}(\bf n),\label{sky}
\end{eqnarray}
where $\bf n$ is a direction in the sky, $X^{\rm CMB}$, ${\tilde A}$ are the underlying CMB signals and foreground contamination, as well as the instrumental noise $N_{\nu}(\bf n)$ at a specific frequency $\nu$. Here, the spectral energy distribution (SED) functions are $a_{\nu}$ and $f_{\nu}$ for the CMB and foreground, respectively. The unit conversion from the brightness temperature to CMB temperature has been done for the SEDs which are normally assumed to be independent of sky directions and can be fully determined by a few foreground parameters. 

In this work, we consider three sources of CMB polarization: (1) the primordial gravitational waves ($X^{\rm PGW}$), (2) the lensing contributions ($X^{\rm lens}$), (3) the primordial magnetic fields ($X^{\rm PMF}$). Thus, the sky map can be described as 
\begin{equation}
X^{\rm CMB}({\bf n})=X^{\rm PGW}({\bf n})+X^{\phi}({\bf n})+X^{\rm PMF}({\bf n}).\label{skycmb}
\end{equation}
The corresponding power spectrum is
\begin{equation}
C_{\ell}^{\rm CMB}=r C_{\ell}^{\rm PGW}+A_{\rm lens}C_{\ell}^{\phi\phi}+A_{\rm PMF}C_{\ell}^{\rm PMF},
\end{equation} where the PMF contains both the tensor and vector contributions, i.e, \begin{equation}
    C_{\ell}^{\rm PMF}=C_{\ell}^{\rm PMF, tens}+C_{\ell}^{\rm PMF, vec},\label{pmfps}
\end{equation}
and $C_{\ell}^{\rm PGW}$ and $C_{\ell}^{\phi\phi}$ are the power spectra of PGW and lensing.

We use the public code \texttt{MagCAMB} to calculate the PMF power spectra~\cite{magcamb}. Here, $r$ is the tensor-to-scalar ratio, $A_{\rm lens}$ is the lensing amplitude and $A_{\rm PMF}$ is the PMF amplitude. The $r\mbox{-}A_{\rm PMF}$ degeneracy is due to the similarity between the PGW and PMF power spectra. However, the PMF can source both the tensor and vector modes with two different power spectra, thus, the degeneracy can be broken by including the vector mode which mainly contributes the power at small scales. We note that the parameter $A_{\rm PMF}$ is proportional to $B^4_{\rm 1Mpc}$ which is the averaged strength of the PMF over the typical 1 Mpc scale. In this work, we fix the epoch of neutrino decoupling which can also affect the power spectrum $C_{\ell}^{\rm PMF, vec}$ and only consider the PMF strength as a free parameter.

Therefore, we can simulate the sky signal in Eq. (\ref{sky}) assuming a future CMB experiment with a specific resolution described by a Gaussian beam profile $b_{\ell}=\exp\{-(\ell(\ell+1)\theta_{\nu})/(16\ln 2)\}$ where the full-width-at-half-maximum (FWHM) is assumed to be frequency dependent as $\theta_{\nu}=\theta_{\rm ref}/\nu\, [{\rm arcmin}]$ with $\theta_{\rm ref}=600$.

Thus, the realistic sky model is 
\begin{eqnarray}
    X^{\rm obs}_{\nu}(\bf n)&=&[a_{\nu}X^{\rm CMB}({\bf n})+f_{\nu}{\tilde A}({\bf n})]\cdot b+N_{\nu}({\bf n}) 
\end{eqnarray}
where $\cdot b$ denotes an operation of beam convolution. Similar to the simulation procedures discussed in the previous works~\cite{momentexp_litebird, planckdiffuse}, we also applied a unit conversion factor between Rayleigh-Jeans brightness temperature and the CMB thermodynamic temperature to the foregrounds. 

If the foreground parameters of the anisotropic SEDs are spatially varying, we can express the dust SED as
\begin{eqnarray}
    f^{(d)}_{\nu}(\bf n)&=&\Big[\frac{\nu}{\nu_0^{(d)}}\Big]^{\beta^{(d)}({\bf n})+1}[e^{\frac{h\nu}{kT^{(d)}(\bf n)}}-1]^{-1}\nonumber\\
    &\sim&\bar f^{(d)}_{\nu}+\Delta\beta^{(d)}({\bf n})\frac{\partial \bar f^{(d)}_{\nu}}{\partial \beta^{(d)}}\Big |_{\beta=\bar\beta^{(d)}}\nonumber\\
    &+&\Delta T^{(d)}({\bf n})\frac{\partial \bar f^{(d)}_{\nu}}{\partial T^{(d)}}\Big |_{T=\bar T}\nonumber\\
    &+&\mathcal{O}([\Delta\beta^{(d)}]^2, [\Delta T^{(d)}]^2),\label{anidustsed}
\end{eqnarray}

and synchrotron SED as
\begin{eqnarray}
    f^{(s)}_{\nu}(\bf n)&=&\Big[\frac{\nu}{\nu_0^{(s)}}\Big]^{\beta^{(s)}({\bf n})}\nonumber\\
    &\sim&\bar f^{(s)}_{\nu}+\Delta\beta_s({\bf n})\frac{\partial \bar f^{(s)}_{\nu}}{\partial \beta^{(s)}}\Big |_{\beta=\bar\beta^{(s)}}\nonumber\\
    &+&\mathcal{O}([\Delta\beta^{(s)}]^2). \label{anisynsed}
\end{eqnarray}
The isotropic SEDs are 
\begin{equation}
    \bar f^{(d)}_{\nu}=\Big[\frac{\nu}{\nu_0^{(d)}}\Big]^{\bar \beta^{(d)}+1}[e^{\frac{h\nu}{k\bar T^{(d)}}}-1]^{-1}\label{seddust}
\end{equation} for dust, and 
\begin{equation}
    \bar f^{(s)}_{\nu}=\Big[\frac{\nu}{\nu_0^{(s)}}\Big]^{\bar \beta^{(s)}}\label{sedsyn}
\end{equation} for synchrotron. In the following text, we assume the SED derivatives are evaluated at the isotropic values and omit the symbols $|_{\beta=\bar\beta^{(s)}}$, $|_{\beta=\bar\beta^{(d)}}$ and $|_{T=\bar T}$. 

If we consider the anisotropic SEDs in Eqs. (\ref{anidustsed}) and (\ref{anisynsed}), the measured fluctuations are 
\begin{eqnarray}
    X^{\rm fg}_{\nu}({\bf n})&=&f^{(c)}_{\nu}({\bf n})\tilde A^{(c)}({\bf n})\nonumber\\
    &\sim&\sum_{c=d,s}\Big\{\bar f^{(c)}_{\nu}\tilde A^{(c)}({\bf n})+\Delta\beta^{(c)}({\bf n})\tilde A^{(c)}({\bf n})\frac{\partial \bar f^{(c)}_{\nu}}{\partial \beta^{(c)}}\nonumber\\
    &+&\frac{1}{2!}[\Delta\beta^{(c)}({\bf n})]^2\tilde A^{(c)}({\bf n})\frac{\partial^2 \bar f^{(c)}_{\nu}}{\partial [\beta^{(c)}]^2}\nonumber\\
    &&+{\rm higher\mbox{-}order}\Big\}\label{FGtaylor},
\end{eqnarray}
where $c=\{d, s\}$ and $X=\{E, B\}$. The spatial components of the anisotropic SEDs will be coupled to the intrinsic intensity fluctuations $\tilde A^{(c)}({\bf n})$, resulting in extra foreground polarization, where $\tilde{A}^{(c)}({\bf n})$ represents the intrinsic amplitude of component $c$ before the spatial SED modulation. 

We assume that the spatial fluctuations of both the intrinsic intensity and the SED spatial fluctuations are Gaussian random fields and the statistical properties are mainly described by the power spectra
\begin{equation}
    \frac{\ell(\ell+1)}{2\pi}C_{\ell}^{(c)}=A_c\Big (\frac{\ell}{\ell_0}\Big )^{\alpha_c}
\end{equation} for the intrinsic intensity and 
\begin{equation}
    C_{\ell}^{\Delta\beta^{(c)}}=B_c\Big (\frac{\ell}{\ell_0}\Big )^{\gamma_c}
\end{equation} for the spatial components of foreground parameters. In this work, we only focus on the non-Gaussian fluctuations generated by the SED variations, and the intrinsic fluctuations $\tilde A^{(c)}({\bf n})$ may also contain non-Gaussian signatures that can be extracted by a quadratic-estimator technique~\cite{ok, hirata}. However, the spatial non-Gaussianity of the foreground intrinsic fluctuation is a different topic from this work, especially, the recent work shows that its impact can be significantly mitigated~\cite{bs}. Therefore, we defer the discussions of the intrinsic non-Gaussian fluctuations to the future work.   

The power spectrum of the foreground including the spatial components is 
\begin{equation}    
C_{\ell}^{{\rm fg},\nu\times\nu'}\delta_{\ell\ell'}\delta_{mm'}=\langle X^{\rm fg}_{\nu, \ell m}X^{\rm fg}_{\nu', \ell' m'}\rangle,
\end{equation}
which is a sum of all the correlations at different perturbation levels. It can be decomposed into
\begin{equation}    
C_{\ell}^{{\rm fg},\nu\times\nu'}\simeq C_{\ell}^{{\rm fg},\nu\times\nu',0\times 0}+ C_{\ell}^{{\rm fg},\nu\times\nu',1\times 1}+ C_{\ell}^{{\rm fg},\nu\times\nu',0\times 2}.\label{fghigherALL}
\end{equation}
Here, the indices ``0", ``1", ``2", which denote the power of $\Delta\beta$ as $(\Delta\beta)^n$, refer to the three terms in Eq. (\ref{FGtaylor}). 

With the assumptions that the spatial components of the foreground parameters are Gaussian, the power spectra of the leading-order perturbations are
\begin{eqnarray}
    C_{\ell}^{{\rm fg},\nu\times\nu',0\times 0}&=&\sum_{c}\bar f^{(c)}_{\nu}\bar f^{(c)}_{\nu'}C_{\ell}^{cc},\nonumber\\
    C_{\ell}^{{\rm fg},\nu\times\nu',1\times 1}&=&\sum_{c}\frac{\partial \bar f^{(c)}_{\nu}}{\partial \beta^{(c)}}\frac{\partial \bar f^{(c)}_{\nu'}}{\partial \beta^{(c)}}\sum_{\ell_1\ell_2}\frac{(2\ell_1+1)(2\ell_2+1)}{4\pi}\nonumber\\
    &\times&\begin{pmatrix}
        \ell&\ell_1&\ell_2\\0&0&0
    \end{pmatrix}^2C_{\ell_1}^{cc}C_{\ell_2}^{\Delta\beta^{(c)}},\nonumber\\
    C_{\ell}^{{\rm fg},\nu\times\nu',0\times 2}&=&\frac{1}{2!}\sum_{c}\Big\{\bar f^{(c)}_{\nu}\frac{\partial^2 \bar f^{(c)}_{\nu'}}{\partial [\beta^{(c)}]^2}+\bar f^{(c)}_{\nu'}\frac{\partial^2 \bar f^{(c)}_{\nu}}{\partial [\beta^{(c)}]^2}\Big\}C_{\ell}^{cc}\nonumber\\
    &\times&\sum_{\ell'}\frac{2\ell'+1}{4\pi}C_{\ell'}^{\Delta\beta^{(c)}}\label{fghiger}.
\end{eqnarray}
Here, the $C_{\ell}^{{\rm fg},\nu\times\nu',0\times 1}$ term is neglected since it is a vanishing bispectrum.  

\begin{figure}
\includegraphics[width=1.0\linewidth]{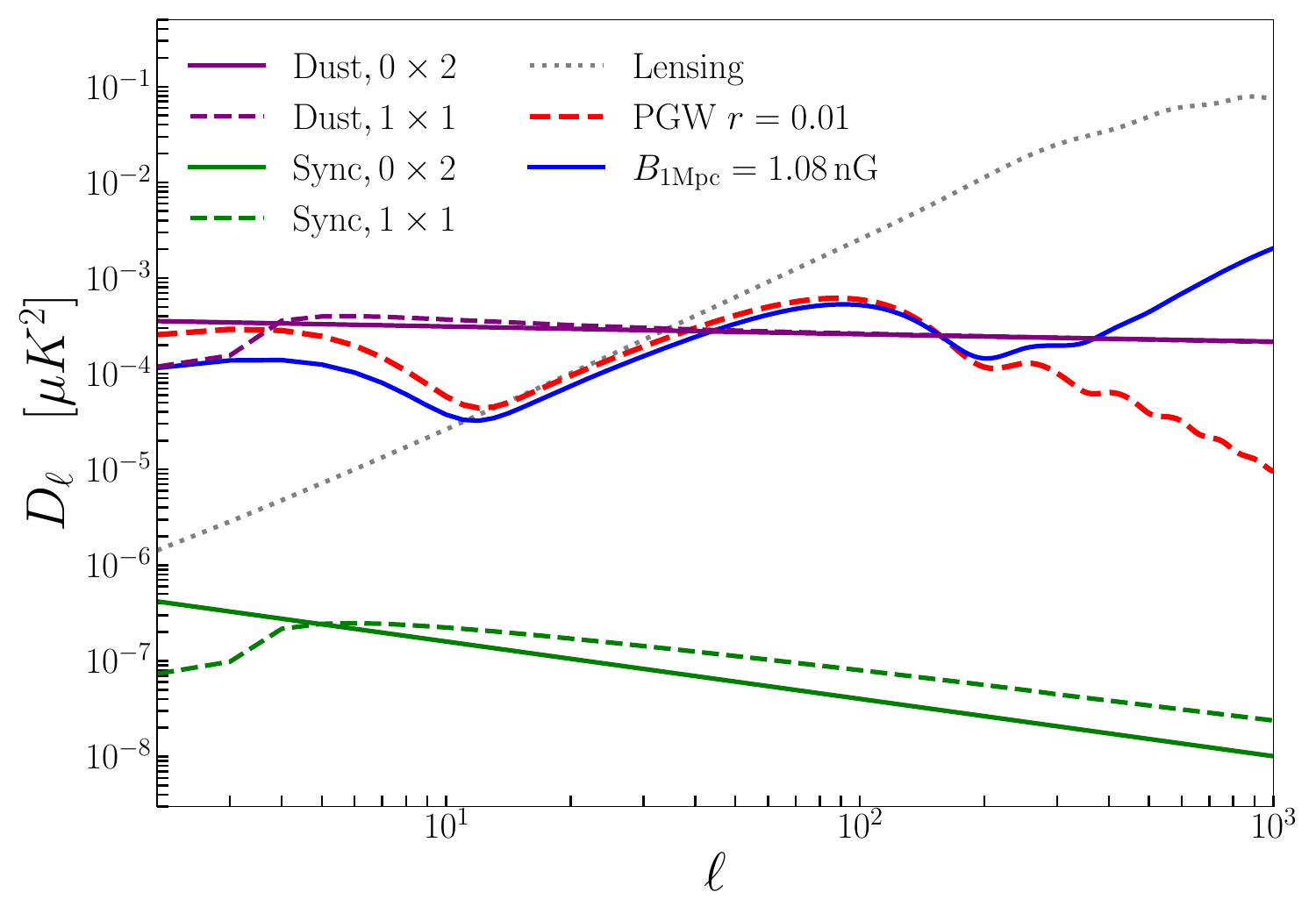}
\caption{A representative plot for different $B$-mode components as described in Eq. (\ref{fghigherALL}). Power spectra of the polarized foregrounds are calculated at 150 GHz. The $B$-mode power spectra of the primordial gravitational waves (PGWs) and the primordial magnetic fields (PMFs) have similar power-spectrum shapes at large angular scales at $\ell<200$. Moreover, the spatial variations of the spectral energy distribution (SED) can also generate higher-order $B$-mode signals with similar descending features. The $y$-axis denotes $D_{\ell}=\ell(\ell+1)/(2\pi)C_{\ell}$. }  \label{degeneracy}
\end{figure}

In Fig. \ref{degeneracy}, we show the detailed $B$-mode components from the PGW, PMF and the SED spatial variations of the polarized foreground at 150 GHz. The PGW and PMF $B$-mode power spectra have similar shapes in power-spectrum space at large angular scales, and the $B$-mode power spectra of the SED spatial variations have similar descending features as the cosmological signatures.

The full parameter set is $\theta=\{A_d, \, \alpha_d,\, B_d, \,\gamma_d,\,A_s, \,\alpha_s, \,B_s,\, \gamma_s\}=\{900{\mu{\rm K}^2},\, -0.16,\, 5 \times 10^{-10},\, -3.5,\, 2000{\mu{\rm K}^2},\, -0.8,\, 2\times 10^{-10},\, -2.5\}$ for the $E$-mode simulation of the polarized foreground, and $\theta=\{A_d, \, \alpha_d,\, B_d, \,\gamma_d,\,A_s, \,\alpha_s, \,B_s,\, \gamma_s\}=\{450{\mu{\rm K}^2},\, -0.08,\, 5 \times 10^{-10},\, -3.5,\, 200{\mu{\rm K}^2},\, -0.6,\, 2\times 10^{-10},\, -2.5\}$ for the $B$-mode simulation of the polarized foreground. The parameters $\gamma_d$ and $\gamma_s$ would cause strong degeneracies so are fixed in this work. The reference frequencies defined in Eqs. (\ref{seddust}, \ref{sedsyn}) are $\nu_0^{(d)}=545$ GHz and $\nu_0^{(s)}=405$ MHz, and the isotropic SED spectral indices are $\bar \beta^{(d)}=1.6$ and $\bar \beta^{(s)}=-3$. The mean dust temperature is $\bar T=21 ${\rm K} and the pivot angular scale $\ell_0=80$. For the cosmological signals, we use the best-fit Planck cosmological parameters~\cite{pk18cosmo}.

\section{Component separation methods in both map- and power-spectrum spaces}
\label{csmethods}

There are two types of solutions to the non-Gaussian foreground. The first approach is to perform a global fitting in one step. The theoretical model for the multifrequency power spectra such as Eq. (\ref{fghigherALL}) can be fitted to the measured ones. For a realistic scenario, a precise model would require more higher-order perturbations to describe the spatial components of the foreground SEDs. However, it would be problematic if many parameters are introduced. Therefore, a truncation of the theoretical model as described in Eq. (\ref{FGtaylor}) though only first- and second-order terms are used and the modeling errors may exist.

The second approach is to extract the three components, i.e, polarized foreground, PGW, and PMF in two steps. A foreground removal procedure can be performed in map space to eliminate the foreground components including the spatial variations. Then the PGW and PMF components can be inferred from the power spectra of component-separated maps. However, the foreground residuals are propagated to the cosmological signals and cause additive biases which may not be negligible, exacerbating the degeneracy issues.

\subsection{Power-spectrum-space methods}

The power-spectrum space methods do not require foreground removal in map space but require precise power-spectrum models for different components.
As mentioned in the previous section, we use the foreground model in Eq. \ref{fghigherALL} with both the first- and second-order perturbations included. The full power spectrum for the sky model in Eq. \ref{sky} is thus
\begin{eqnarray}
    C_{\ell}^{\nu\times\nu'}&=&C_{\ell}^{{\rm fg},\nu\times\nu',0\times 0}+ C_{\ell}^{{\rm fg},\nu\times\nu',1\times 1}+ C_{\ell}^{{\rm fg},\nu\times\nu',0\times 2}\nonumber\\
    &&+rC_{\ell}^{\rm PGW}+A_{\rm lens}C_{\ell}^{\rm lens}+A_{\rm PMF}(C_{\ell}^{\rm PMF, tens}\nonumber\\
    &&+C_{\ell}^{\rm PMF, vec})+N_{\ell}^{\nu\times\nu'}.\label{skyps}
\end{eqnarray}

The covariance of the multifrequency power spectra can be described by the Knox formula as
\begin{eqnarray}
    \mathcal{C}(x^{\nu_1\times\nu_2}, x^{\nu_3\times\nu_4})&=&\frac{1}{(2\ell+1)f_{\rm sky}\Delta\ell}\Big[x^{\nu_1\times\nu_3}x^{\nu_2\times\nu_4}\nonumber\\
    &&+x^{\nu_1\times\nu_4}x^{\nu_2\times\nu_3}\Big].
\end{eqnarray}

We perform a global fitting using the Bayesian analysis and the likelihood function is
\begin{equation}
    -\ln{\mathcal{L}}=\frac{1}{2}[\hat {\bf x}-{\bf x}(\theta)]\mathcal{C}^{-1}[\hat {\bf x}-{\bf x}(\theta)]+\frac{1}{2}\ln{{\rm det}\mathcal{C}},\label{bayesian}
\end{equation}
where $\theta$ denotes all the cosmological and astrophysical parameters, ${\bf x}=\{x^{\nu_1\times\nu_2}\}$ denotes a power spectrum vector at any two frequencies $\nu_1$ and $\nu_2$.

\begin{figure*}
\includegraphics[width=0.8\linewidth]{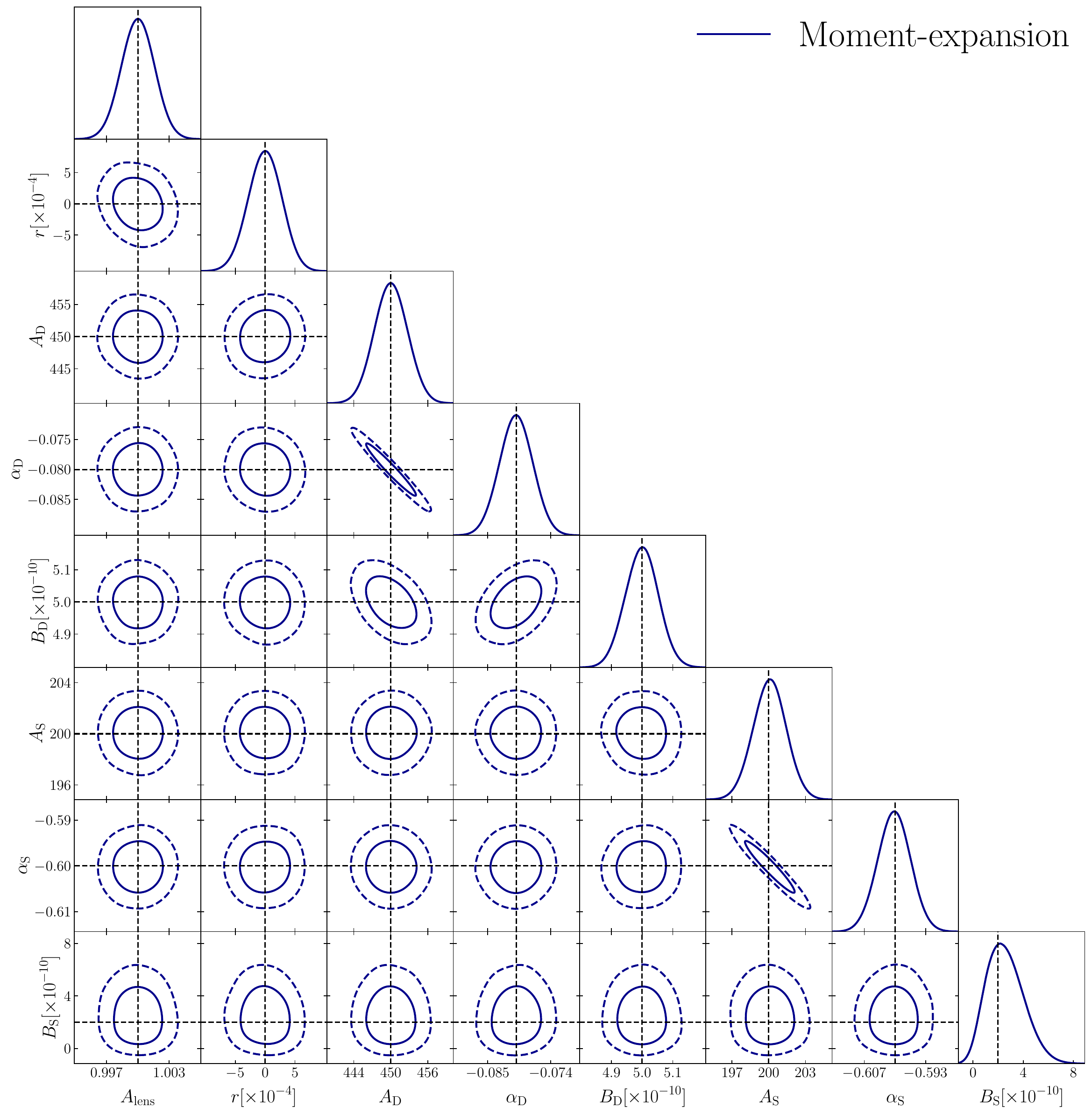}
\caption{Posterior distribution functions of CMB and foreground parameters. The mock power spectra (mock data 2 in Table \ref{simtype})) are used for the Bayesian analysis described in Eq. (\ref{bayesian}). The two component CMB model $r+A_{\rm lens}$ is inferred in the presence of polarized foreground using the power-spectrum-based method.} \label{mapcheck1}
\end{figure*}

We create mock power spectra using the power-spectrum model in Eq. (\ref{skyps}) with the fiducial cosmological parameters as $\textbf{P}=\{r, A_{\rm lens}, A_{\rm PMF}\}=\{0, 1, 0\}$. This mock data set is referred to as \texttt{power-spectrum moment-expansion} which is listed in Table \ref{simtype}.   In Fig. \ref{mapcheck1}, we test the power-spectrum model with two free cosmological parameters $r$ and $A_{\rm lens}$ and run the Markov chain Monte Carlo (MCMC) analysis as described in Eq. (\ref{bayesian}). The priors on the cosmological and foreground parameters are listed in Table \ref{priors}. The posterior distribution functions (PDFs) from the MCMC samples consistently peak around the input values of both the foreground and CMB parameters and the tensor-to-scalar ratio $r$ is not correlated with the foreground parameters. 
 
\begin{figure*}
\includegraphics[width=0.8\linewidth]{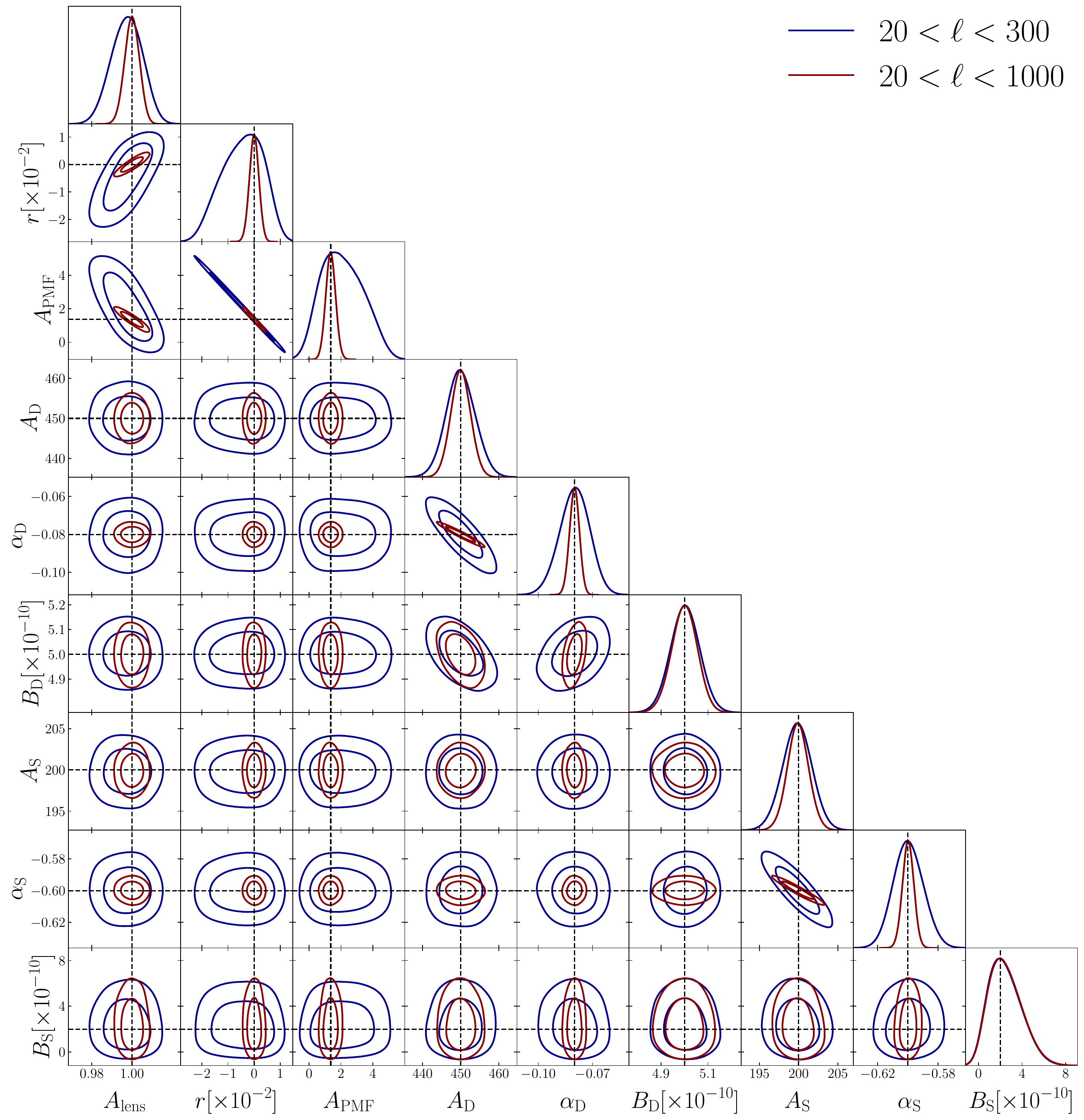}
\caption{Posterior distributions functions of foreground and cosmological parameters. This test shows the impact of different multipole ranges.} \label{compare_r}
\end{figure*}

Next, we consider the model with $\{r, A_{\rm lens}, A_{\rm PMF}\}=\{0, 1, 1.08^4\}$ where the PMF strength is fixed at a representative strength $B_{\rm 1Mpc}=1.08$ nG. Fig. \ref{compare_r} shows the PDFs for all parameters consisting of the CMB parameters $r+A_{\rm lens}+A_{\rm PMF}$ and the foreground parameters. As investigated in~\cite{degeneracy, degeneracy2}, there is a strong degeneracy between $r$ and $A_{\rm PMF}$ due to the similar power spectrum shapes of the $C_{\ell}^{\rm PGW}$ and $C_{\ell}^{\rm PMF, tens}$. So the PMF-induced $B$-mode signal may leak into the PGW $B$-mode signals. Also, there are mild correlations between $r$ and the foreground parameters when the parameter $A_{\rm PMF}$ is also varying. The parameter degeneracies can be significantly reduced by extending to higher multipoles. This is demonstrated in Fig. \ref{compare_r} with two sets of contours corresponding to $\ell_{\rm max}=300$ and $\ell_{\rm max}=1000$. 

The $\ell_{\rm min}$ also has nonnegligible impact on the parameter constraints and a lower $\ell_{\rm min}$ can effectively reduce the errors on the cosmological parameters. However, as found in~\cite{cmilc}, the foreground residuals from different component separation methods could be much higher than the noise biases in the multipole range $\ell<15$, potentially introducing certain biases to the inferred parameters. As a conservative approach, we chose $\ell_{\rm min}=20$ for the map-based method.

\begin{table}[ht]
\centering
\small
\caption{Different mock datasets investigated in this work.}
\label{simtype}
\begin{tabular}{llll}
\toprule
  &\textbf{Mock data} & \textbf{Foreground} & \textbf{Type} \\
\midrule
1&\texttt{map moment-expansion}& Eq. (10) & baseline\\
2 & \makecell[tl]{\texttt{power-spectrum}\\\texttt{moment-expansion}} & Eq.~(14) & baseline \\
3&\texttt{PySM simulation}&-& validation-only\\
\bottomrule
\end{tabular}
\end{table}

\begin{table}[htbp]
\centering
\caption{Uniform priors adopted in the MCMC constraints in Figs. \ref{mapcheck1} and \ref{compare_r} for the power-spectrum-based method.}
\label{priors}
\begin{tabular}{ll}
\toprule
\textbf{Parameter} &  \textbf{Priors} \\
\midrule
$A_{\text{lens}}$ &  $\mathcal{N}[0,\,2]$ \\
$r$               &  $\mathcal{N}[-0.5,\,0.5]$ \\
$A_{\text{PMF}}$  & $\mathcal{N}[0,\,3]$ \\
$A_D$             & $\mathcal{N}[420,\,480]$ \\
$\alpha_D$        & $\mathcal{N}[-0.2,\,0]$ \\
$B_D$             & $\mathcal{N}[0,\,2\times10^{-9}]$ \\
$A_S$             & $\mathcal{N}[180,\,22 0]$ \\
$\alpha_S$        & $\mathcal{N}[-0.7,\,-0.5]$ \\
$B_S$             & $\mathcal{N}[0,\,10^{-9}]$ \\
\bottomrule
\end{tabular}
\end{table}

\subsection{Map-space methods}
The map-space methods employ a weighting function $\bf w$ in either map-space or harmonic space, such as 
\begin{equation}
    \hat s={\bf w}^T{\bf X}^{\rm obs}\label{cs}
\end{equation}
to extract the desired signals from the  multifrequency observations ${\bf X}^{\rm obs}$. Here, the superscript ``$T$" denotes a matrix transpose operation. In order to require that the estimated signal $\hat s$ is both unbiased and has minimum variance, we can optimally choose the weighting function which can extract the CMB signal while eliminating other foreground terms including the higher-order perturbations. These constraints can be imposed on the weighting functions as 
\begin{equation}
    {\bf w}^T{\bf \Lambda}= {\bf V}^T
\end{equation}
where the matrix ${\bf \Lambda}$ is
\begin{equation}
    {\bf \Lambda}=\Big[{\bf a},\,{\bf \bar f}^{(d)},\, \frac{\partial {\bf \bar f}^{(d)}}{\partial \beta^{(d)}}, \,{\bf \bar f}^{(s)}, \frac{\partial {\bf \bar f}^{(s)}}{\partial \beta^{(s)}} \Big]\label{matlambda}
\end{equation}
and ${\bf V}$ is
\begin{equation}
    {\bf V}^T=\Big[1, 0,0,0,0\Big].\label{matv}
\end{equation}

Thus, the pixel variance of the estimated signal $\hat s$ is
\begin{equation}
    \sigma_s^2=\langle\hat{ s}\hat{s}^T\rangle={\bf w}^T\mathcal{C}^{\rm obs}{\bf w},
\end{equation}
where the covariance matrix of the observed CMB signals is $\langle\hat{\bf X}^{\rm obs}(\hat{\bf X}^{\rm obs})^T\rangle=\mathcal{C}^{\rm obs}$.

To achieve the minimum variance while obtaining an unbiased signal, we can rewrite the pixel variance as a functional form with a Lagrange multiplier $\bf \Gamma$ as 
\begin{eqnarray}
    (\sigma')^2_s&=&\sigma_s^2+{\bf \Gamma}^T[{\bf V}-{\bf \Lambda}^T{\bf w}]\nonumber\\
    &=&{\bf w}^T\mathcal{C}{\bf w}+{\bf \Gamma}^T[{\bf V}-{\bf \Lambda}^T{\bf w}].
\end{eqnarray}
When the functional derivative is zero, i.e., 
$\frac{\delta (\sigma')^2_s}{\delta {\bf w}}=0$, the optimal weighting is determined and is
\begin{equation}
    {\bf w}^T={\bf V}^T({\bf \Lambda}^T\mathcal{C}^{-1}{\bf \Lambda})^{-1}{\bf \Lambda}^T\mathcal{C}^{-1}.\label{opw}
\end{equation}

To robustly estimate the covariance matrix of the observed CMB maps, we adopt the needlet ILC scheme with ten cosine-like needlets $h_{\ell}^{(c)}$~\cite{nilc} and decompose the observed CMB maps into different angular scales with indices $(j)$, i.e, $X^{\rm obs}_{\nu}({\bf n})=\sum_{j}X^{\rm obs, (j)}_{\nu}({\bf n})$. Then we perform the ILC for the map $X^{\rm obs,(j)}_{\nu}({\bf n})$ at each needlet scale. Here, the scale index is $j\in \{1,10\}$. 

The covariance matrix in Eq. (\ref{opw}) is calculated for each scale, following the definition 
\begin{equation}
    \mathcal{C}=\frac{1}{\mathcal{N}}\sum_{k\in \mathcal{D}_k} \hat{\bf X}^{\rm obs}({n_k})(\hat{\bf X}^{\rm obs}({n_k}))^T.
\end{equation}
The summation is done within a region $\mathcal{D}_k$ centered around the pixel $k$ and $\mathcal{N}$ is the total number of pixels in the region. The choice of the region $\mathcal{D}_k$ is not trivial and can introduce the ILC bias~\cite{ilcbias}. In this work, we use the same implementation as~\cite{pyilc} and determine the size of the region by requiring the ILC bias to be 1\%.

By choosing different elements in matrix $\bf \Lambda$ (Eq. (\ref{matlambda})), we can construct different component separation methods. For example, the method described by the matrix ${\bf \Lambda}=\Big[{\bf a},\,{\bf \bar f}^{(d)}, \,{\bf \bar f}^{(s)}\Big]$ corresponds to the constrained ILC (cILC), $\Big[{\bf a},\,{\bf \bar f}^{(d)},\, \frac{\partial {\bf \bar f}^{(d)}}{\partial \beta^{(d)}}, \,{\bf \bar f}^{(s)}\Big]$ corresponds to cMILC06, and $\Big[{\bf a},\,{\bf \bar f}^{(d)},\, \frac{\partial {\bf \bar f}^{(d)}}{\partial \beta^{(d)}}, \,{\bf \bar f}^{(s)},\,\frac{\partial {\bf \bar f}^{(s)}}{\partial \beta^{(s)}}\Big]$ corresponds to cMILC07. If the SED derivative with respect to the first-order dust temperature is also included in ${\bf \Lambda}$ of cMILC07, it is extended to cMILC08. The definition of these methods is the same as those in~\cite{cmilc}.

\subsubsection{Mock data and power-spectrum estimation for the map-space method}
\label{results}

In this work, we use an end-to-end foreground modeling as described in Eq. (\ref{FGtaylor}), so in principle we can generate the spatial components of the foreground SEDs at any perturbation levels and can theoretically calculate the power spectra among different perturbations. This mock dataset is referred to as \texttt{map moment-expansion} as listed in Table \ref{simtype}. These end-to-end simulations allow precise comparisons of different foreground removal methods in both the map space and power-spectrum space since all the input parameters are known. We note that only the first-order spatial components are considered for tests in map space because we only consider first few moments such as cMILC03 and cMILC07 for simplicity and the second-order perturbations in the map space would require more moments than those included in cMILC08. However, both the first- and second-order spatial components are included for tests with the power-spectrum-based method. 

We consider a futuristic full-sky CMB experiment with a white noise level at $1\mu{\rm K}\mbox{-}{\rm arcmin}$ but a frequency-dependent FWHM as described in Sec. \ref{skysec}. We consider seven frequency bands between $20\le\nu\le350$, i.e., $\nu_b\in\{20, 50, 100, 150, 250, 300, 350\}$ GHz. Specifically, we generate Gaussian CMB realizations for the maps described in Eq. (\ref{skycmb}) and the polarized foreground maps using Eq. (\ref{FGtaylor}) where only the first-order spatial components of the spectral indices are considered. Therefore, the higher-order terms $\sum_{c=d,s}\Delta\beta^{(c)}({\bf n}) \tilde A^{(c)}({\bf n})$ would be the source for the foreground residuals if the standard foreground removal procedures were adopted.

\begin{figure*}
\includegraphics[width=0.9\linewidth]{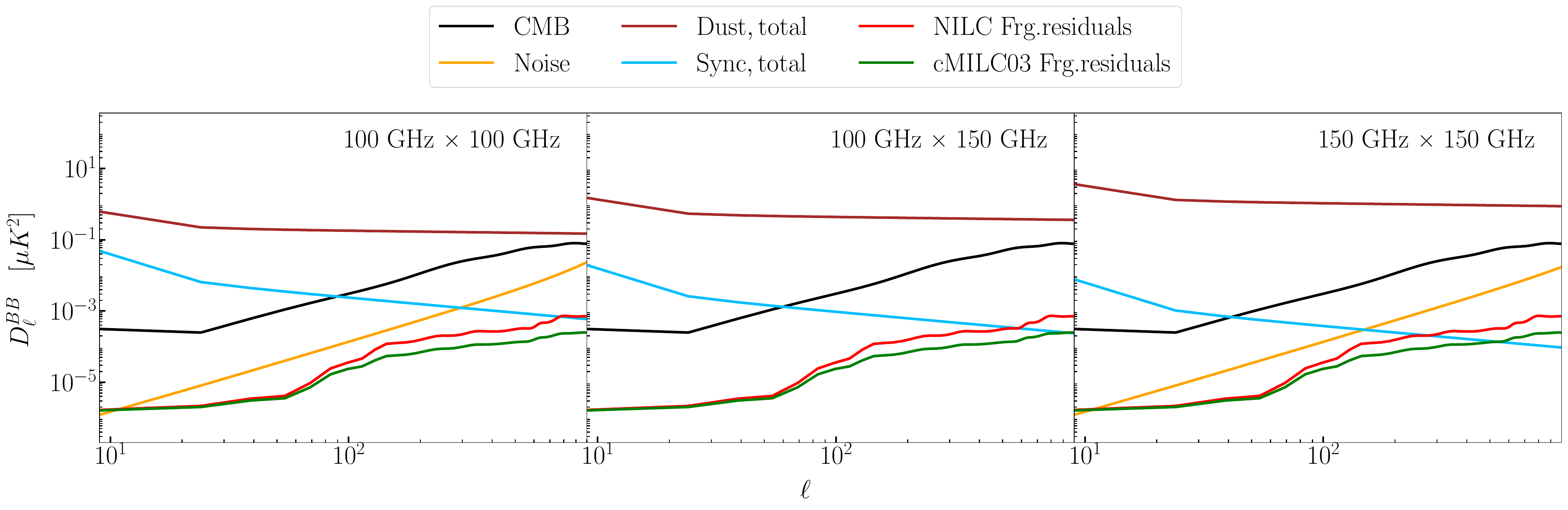}
\caption{Power-spectrum components described in Eq. (\ref{fghigherALL}). The power spectra are calculated at seven frequencies, i.e., $\nu$=\{20, 50, 100, 150, 250, 300, 350\} GHz. The components labeled with ``total" refer to all the power-spectrum components defined in Eq. (\ref{fghigherALL}).}  \label{allpsmock}
\end{figure*}

Specifically, we show the theoretical power spectra at different perturbation levels in Fig. \ref{allpsmock}. For the seven frequency bands between $20\le \nu \le 350$ GHz, the higher-order contributions are not negligible.

We create both the ``raw map" dataset which contains all the components in Eq. (\ref{sky}) and ``noise map" dataset which only contains the noise components. 1000 realizations are generated for each dataset with experimental settings incorporated. We apply the component separation procedures in Eq. (\ref{cs}) to two suites of mock data and obtain ensembles of $\{\hat s^{\rm raw\,map}\}$ and $\{\hat s^{\rm noise\,map}\}$. The power spectra of the component-separated maps from the ``raw map" are 
\begin{equation}
    \hat C_{\ell}^{\rm raw\,map}\delta_{\ell\ell'}\delta_{mm'}=\langle\hat s^{\rm raw\,map}_{\ell m}\hat s^{\rm raw\,map, \ast}_{\ell' m'}\rangle,
\end{equation}
and the ones from ``noise map" are
\begin{equation}
    \hat C_{\ell}^{\rm noise\,map}\delta_{\ell\ell'}\delta_{mm'}=\langle\hat s^{\rm noise\,map}_{\ell m}\hat s^{\rm noise\,map, \ast}_{\ell' m'}\rangle.
\end{equation}
Both power spectra are calculated by \texttt{NaMaster} with a binning size $\Delta\ell=15$.

The estimated signal is thus
\begin{equation}
    \hat C_{\ell}^{\rm est}=\hat C_{\ell}^{\rm raw\,map}-\hat C_{\ell}^{\rm noise\,map}-A_{\rm lens}C_{\ell}^{\rm lensing}.\label{estcl}
\end{equation}
Here, $A_{\rm lens}$ refers to the amount of lensing assumed in the mock data. In this work, we just assume a lensing residual as quantified by $A_{\rm lens}C_{\ell}^{\rm lensing}$ where $C_{\ell}^{\rm lensing}$ is a template. However, the delensing procedures~\cite{blakecibdelens15} would require a split of the multipole range into two regions for both the PGW inference and lensing reconstruction. However, for the future multifrequency and high-sensitivity CMB datasets, the Bayesian delensing approaches, such as the joint estimation of the lensing potential and the delensed CMB maps~\cite{bayesiandelensing17,bayesiandelens} and the maximum a priori approach~\cite{mapdelensing}, can avoid such a $\ell$-split and simultaneously infer the blended cosmological signals including the tensor-to-scalar ratio $r$ and the PMF strength $A_{\rm PMF}$ while delensing the CMB maps. The estimated signal is a sum of cosmological signals and foreground residuals. The band-power errors are theoretically calculated using the Knox formula
\begin{eqnarray}
    \Delta C_{\ell}^{BB}&=&\sqrt{\frac{2}{(2\ell+1)f_{\rm sky}\Delta\ell}}[rC_{\ell}^{\rm PGW}+A_{\rm lens}C_{\ell}^{\rm lensing}\nonumber\\
    &&+{\hat N}^{BB}_{\ell}+\Delta C_{\ell}^{\rm fg}],\label{bberr_knox}
\end{eqnarray}
but we also estimated the band powers from the ensemble of $B$-mode power spectrum and the simulated band-power errors are consistent with Eq. (\ref{bberr_knox}). Here, $f_{\rm sky}$ is the sky fraction, ${\hat N}^{BB}_{\ell}$ is the ensemble average of $\{{\hat C}^{\rm noise\,map}_{\ell}\}$ and $\Delta C_{\ell}^{\rm fg}$ is the average of $\{\hat C_{\ell}^{\rm est}\}$. The last term is the ensemble average for the foreground residuals described in Eq. (\ref{estcl}). 

By comparing the cosmological models with the residuals described in Eq. (\ref{estcl}), we infer different parameters $\textbf{P}=\{r, A_{\rm lens}, A_{\rm PMF}\}$ from the likelihood function
\begin{eqnarray}
    &&-2\log\mathcal{L}(\textbf{P})=\sum_{\ell}\Big[\frac{\hat C_{\ell}^{\rm est}-C_{\ell}^{\rm model}(\textbf{P})}{\Delta C_{\ell}^{BB}}\Big]^2.
\end{eqnarray} where the model is $C_{\ell}^{\rm model}(\textbf{P})=rC_{\ell}^{\rm PGW}+A_{\rm lens}C_{\ell}^{\rm lensing}+A_{\rm PMF}C_{\ell}^{\rm PMF}$.

\subsubsection{Results with map-space methods}
\label{mapspacetests}

\begin{figure*}
\includegraphics[width=0.9\linewidth]{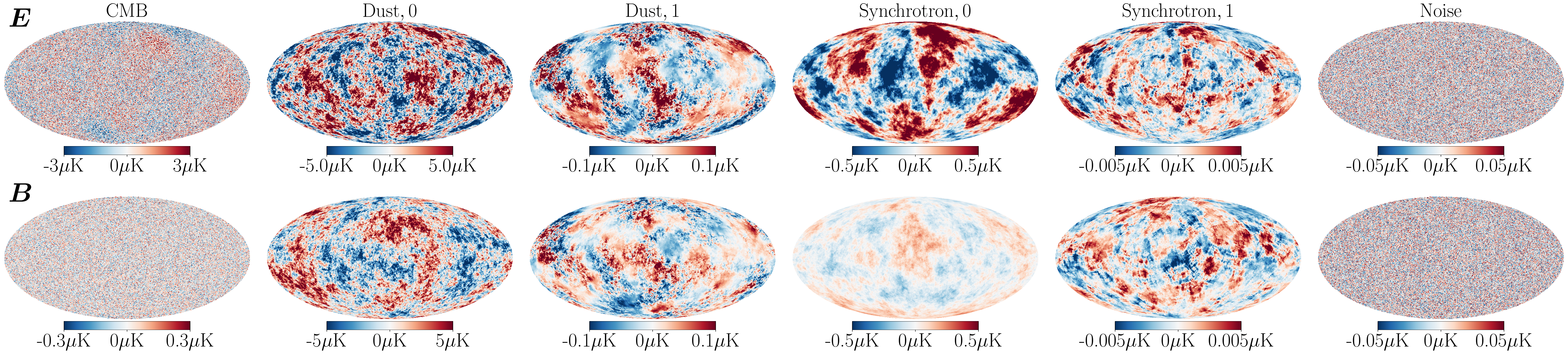}
\caption{Different $E$-mode and $B$-mode components at 150 GHz. Maps are filtered by a top-hat window at $2\leq l\leq 300$ in Fourier space to visualize the large-scale modes. } \label{skymapEB}
\end{figure*}

In Fig. \ref{skymapEB}, we show the simulated polarization maps for CMB and polarized foreground, as well as the noise realizations at a representative frequency 150 GHz. The dust polarization $E^{\rm Dust, 0}$ or $B^{\rm Dust, 0}$ dominate the sky polarization signals as indicated by the color scales. The higher-order perturbations are generated from two independent Gaussian realizations as described in Eq. (\ref{FGtaylor}). The maps are filtered with a top-hat window at $2<\ell<300$ to highlight the large-scale fluctuations.

\begin{figure*}
\includegraphics[width=0.9\linewidth]{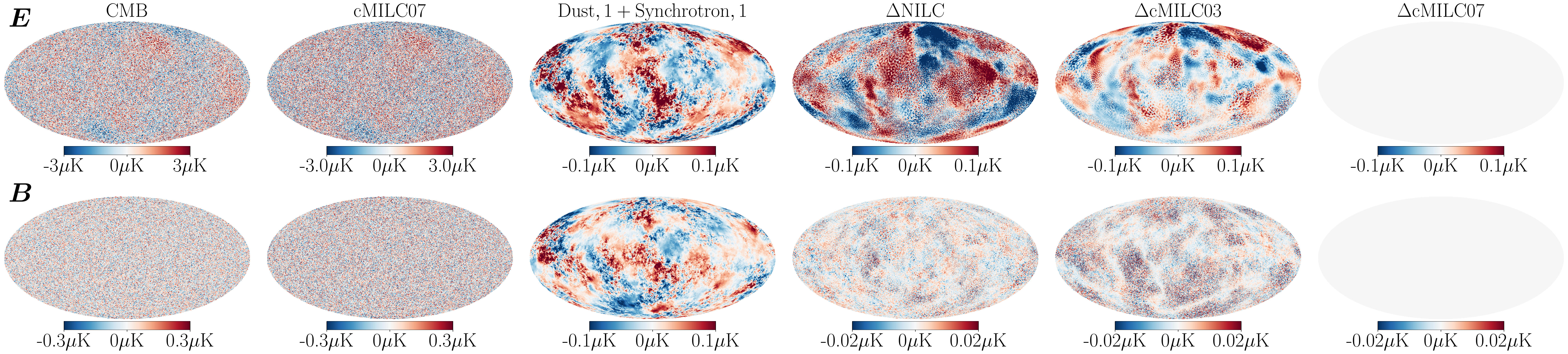}
\caption{Comparison of reconstructed maps using different foreground removal methods. The first and second columns show the input and reconstructed $E$- and $B$-mode maps. The last three columns labeled by a $\Delta$ in front of the method show the foreground residuals from different methods in map space, as well as the expected overall first-order spatial contributions in the third column. The last column for the cMILC07 scheme contains no visible fluctuations, demonstrating that the cMILC07 algorithm can successfully remove the spatial components of the polarized foregrounds.} \label{skymapaftercsEB}
\end{figure*}

In Fig. \ref{skymapaftercsEB}, we show the reconstructed polarization maps $\hat s^{\rm raw\,map}$ after implementing different component-separation procedures. The first column is the input polarization signal and the second column shows the component-separated polarization maps which are consistent with the input ones. The third column is the expected foreground residuals due to the SED variations and we compare foreground residuals from foreground removal procedures, i.e., NILC, cMILC03 and cMILC07 in the last three columns. The foreground residuals are gradually reduced when more moments (the elements in matrix $\bf \Lambda$) are included in the weighting functions (Eq.(\ref{opw})). The cMILC07 which is constructed to eliminate the first-order perturbations of the foreground spectral indices is proven to be able to remove these contributions completely as seen from the last column. 

We perform different tests which are summarized in Table \ref{testtype} to investigate the impact of the non-Gaussian foreground residuals on the cosmological signals. To further quantify the foreground residuals in the reconstructed maps shown in Fig. \ref{skymapaftercsEB}, we calculate the power spectra of the reconstructed maps with \texttt{NaMaster} for two scenarios. 
We first assume there is no inflationary $B$-mode signal ($r=0$) in the ``raw map" mock data (test I in Table \ref{testtype}). 

\begin{figure*}
\includegraphics[width=0.9\linewidth]{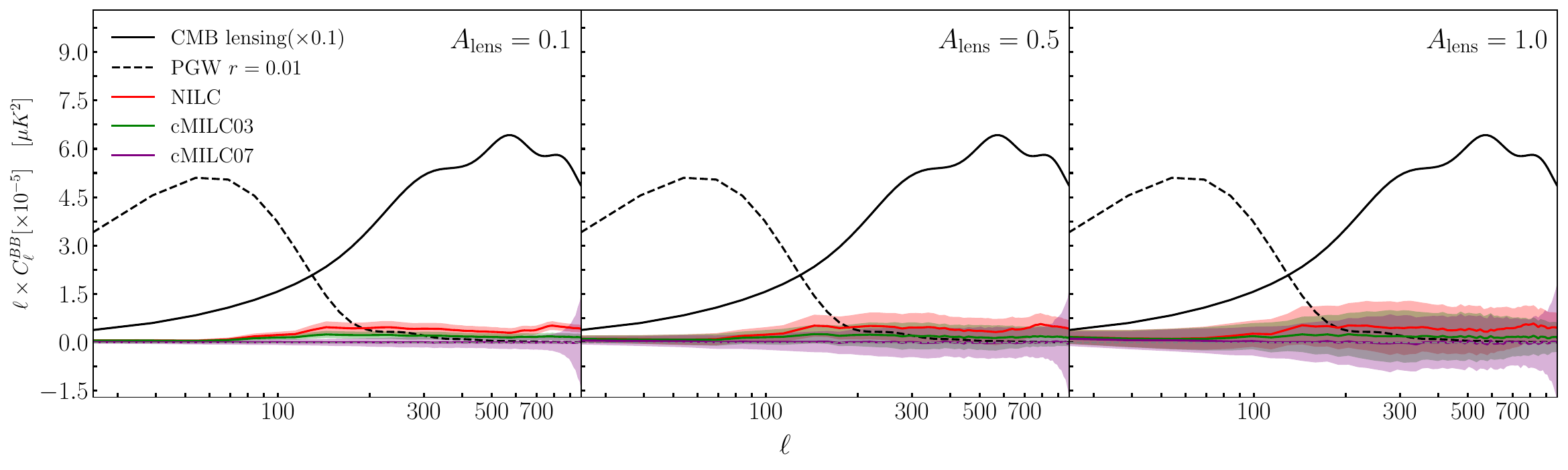}
\caption{Component-separated $B$-mode power-spectra from the mock data simulated at seven frequency bands. CMB lensing $B$-mode signals are simulated with different amplitudes: $A_{\rm lens}=0.1$ (left), $A_{\rm lens}=0.5$ (middle), and $A_{\rm lens}=1$ (right). There is no inflationary $B$-mode signal ($r=0$) included in the mock data. The cMILC07 method yields the smallest foreground residuals.}\label{bbnull}
\end{figure*}

\begin{figure*}
\includegraphics[width=0.9\linewidth]{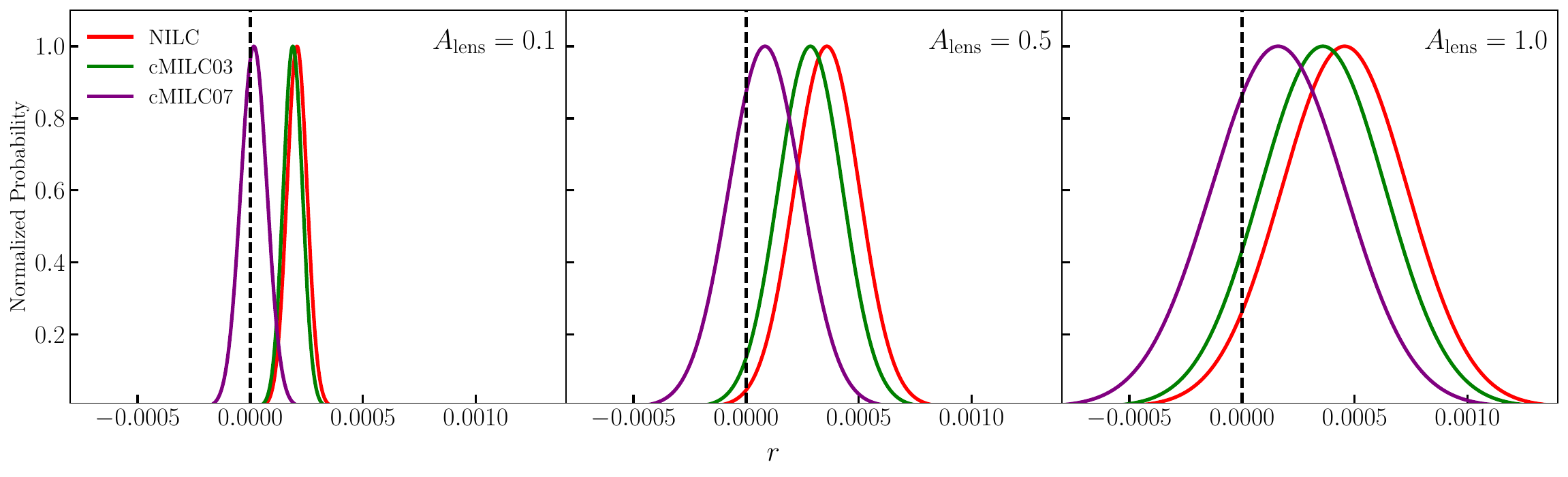}
\caption{Posterior distribution functions (PDFs) of the tensor-to-scalar ratio $r$ with different foreground removal methods. The comparison of the PDFs is made for three different lensing $B$-mode amplitudes introduced in Fig. \ref{bbnull}. The methods with more moments which can eliminate more non-Gaussian structures in the foreground SEDs can significantly reduce the foreground residuals but will increase the errors as a noise penalty.} \label{like_bbnull}
\end{figure*}

\begin{figure}
\includegraphics[width=8cm, height=5cm]{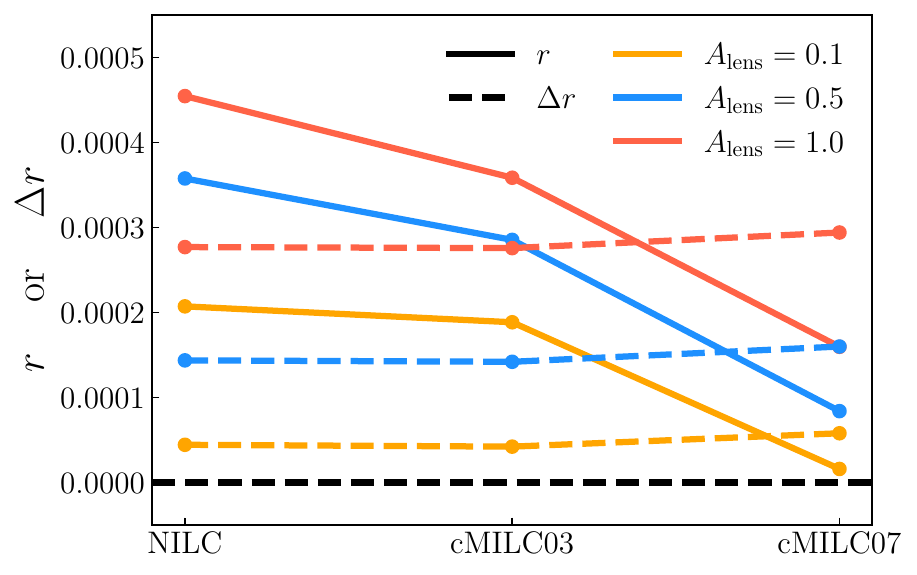}
\caption{A summary of the biases and errors on the tensor-to-scalar ratio $r$ illustrated in Fig. \ref{like_bbnull}. } \label{stat_bbnull}
\end{figure}

In Fig. \ref{bbnull}, we compare the results from the three foreground removal procedures, i.e, NILC, cMILC03 and cMILC07. Each foreground residual labeled by the method name is derived from Eq. (\ref{estcl}). The power spectra of the foreground residuals derived from cMILC07 are vanishing whereas residual power spectra of NILC and cMILC03 are still nonnegligible because of the spatial fluctuations in the spectral indices $\beta^{(d)}(\bf n)$ and $\beta^{(s)}(\bf n)$.
We also investigate the impact of lensing $B$-mode signals on the residual power spectra. Specifically, we consider three lensing amplitudes, i.e, $A_{\rm lens}=\{1, 0.5, 0.1\}$ and add Gaussian realizations of lensing $B$-mode signals with different amplitudes into the mock data. From Figs. \ref{like_bbnull} and \ref{stat_bbnull}, it is seen that the cMILC07 yields much smaller foreground residuals but larger errors on $r$ than the rest with less moments. This is the expected tradeoff for achieving lower biases while sacrificing the precision as discussed in~\cite{cmilc}. Also, a small amount of lensing signals can effectively reduce the errors on $r$ as the posterior distribution functions (PDFs) for a smaller $A_{\rm lens}$ become much sharper. 


\begin{figure*}
\includegraphics[width=0.325\linewidth]{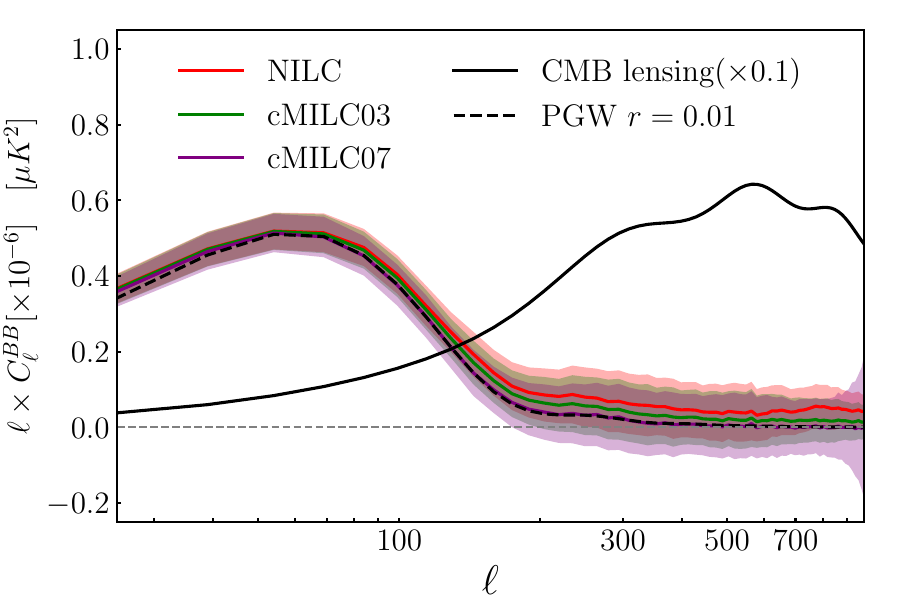}
\includegraphics[width=0.326\linewidth]{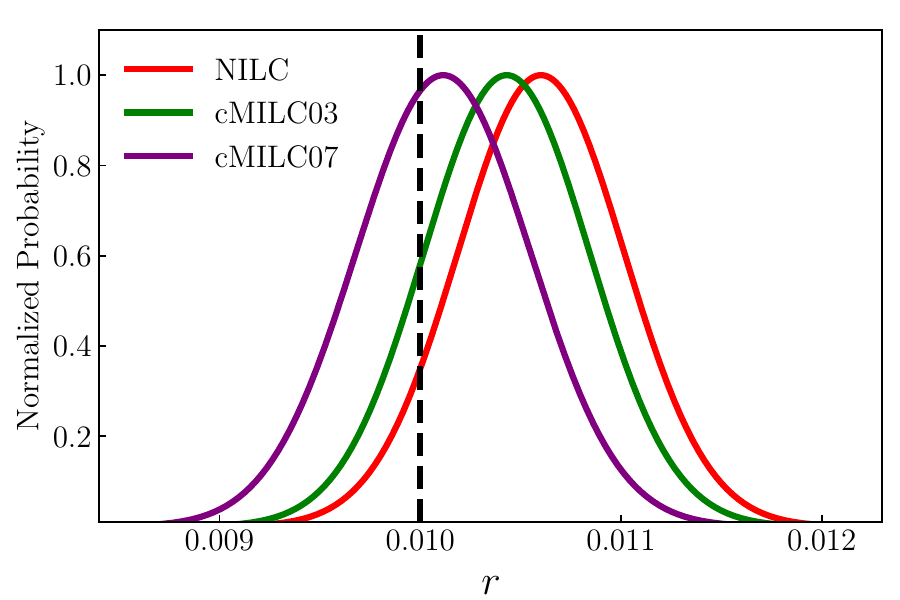}
\includegraphics[width=0.326\linewidth]{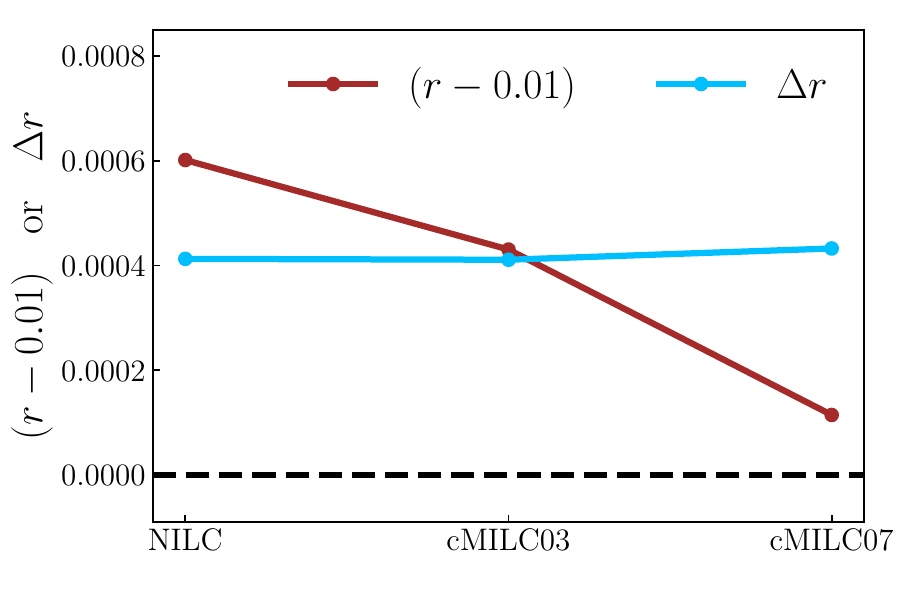}
\caption{Foreground removal when there is a PGW signal at $r=0.01$. The left subplot is the reconstructed $B$-mode power spectrum, the middle is the PDF of $r$ from each foreground removal method, and the last is the bias and error comparison. }\label{bbsignal}
\end{figure*}

The previous test in Fig. \ref{bbnull} does not include the inflationary $B$-mode signals in the ``raw map" so the estimated signals in Eq. (\ref{estcl}) are just the foreground residuals with the instrumental noise. Next, we add the inflationary $B$-mode signal with $r=0.01$ in the ``raw map" and repeat the same analysis as Fig. \ref{bbnull} (test II in Table \ref{testtype}). As seen from Fig. \ref{bbsignal}, the estimated signals from three methods are all consistent with the input inflationary $B$-mode signal below $\ell<100$ but start to deviate from the theory beyond $\ell>100$ except the cMILC07. 

\begin{table}[t]
\centering
\caption{Different test combinations of fiducial models and free parameters used in Sec. \ref{mapspacetests}.}
\label{testtype}
\scalebox{0.8}{
\begin{tabular}{llll}
\toprule
  &\textbf{Fiducial models} & \textbf{Free parameters}  \\
\midrule

I & $A_{\rm lens}=(1,0.5,0.1), A_{\rm PMF}=0,r=0$&$r$&Fig. \ref{bbnull}\\
II & $A_{\rm lens}=1, A_{\rm PMF}=0,r=0.01$&$r$&Fig. \ref{bbsignal}\\
III & $A_{\rm lens}=1, A_{\rm PMF}=0,r=0$&$r+A_{\rm lens}$&Fig. \ref{mapcheck2}\\
IV & $A_{\rm lens}=1, A_{\rm PMF}=1.08^4,r=0$&$r+A_{\rm PMF}$&Fig. \ref{threesignals}\\
V& $A_{\rm lens}=1, A_{\rm PMF}=1.08^4,r=0$&$r+A_{\rm lens}+A_{\rm PMF}$&Fig. \ref{three_component_comparison}\\
\bottomrule
\end{tabular}}
\end{table}

\begin{figure}
\includegraphics[width=0.8\linewidth]{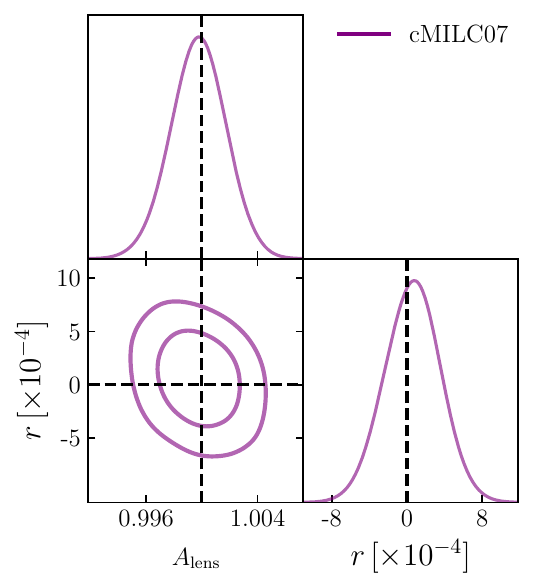}
\caption{Tests of the two-component CMB model $r$+$A_{\rm lens}$ with non-Gaussian polarized foregrounds (test III in Table \ref{testtype}). The multipole range for this analysis is $20<\ell<1000$.} \label{mapcheck2}
\end{figure}

\begin{figure}
\includegraphics[width=8cm, height=8cm]{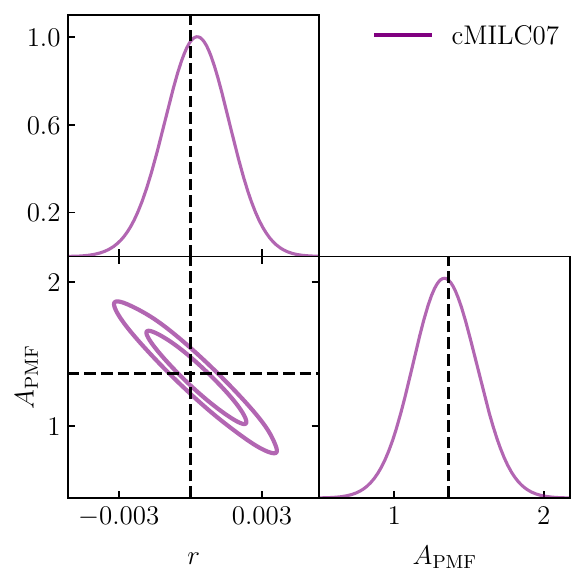}
\caption{Tests of the three-component CMB model $r$+$A_{\rm lens}$+$A_{\rm PMF}$ with non-Gaussian polarized foregrounds (test IV in Table \ref{testtype}). The multipole range for this analysis is $20<\ell<1000$.} \label{threesignals}
\end{figure}

\begin{figure}
\includegraphics[width=8cm, height=8cm]{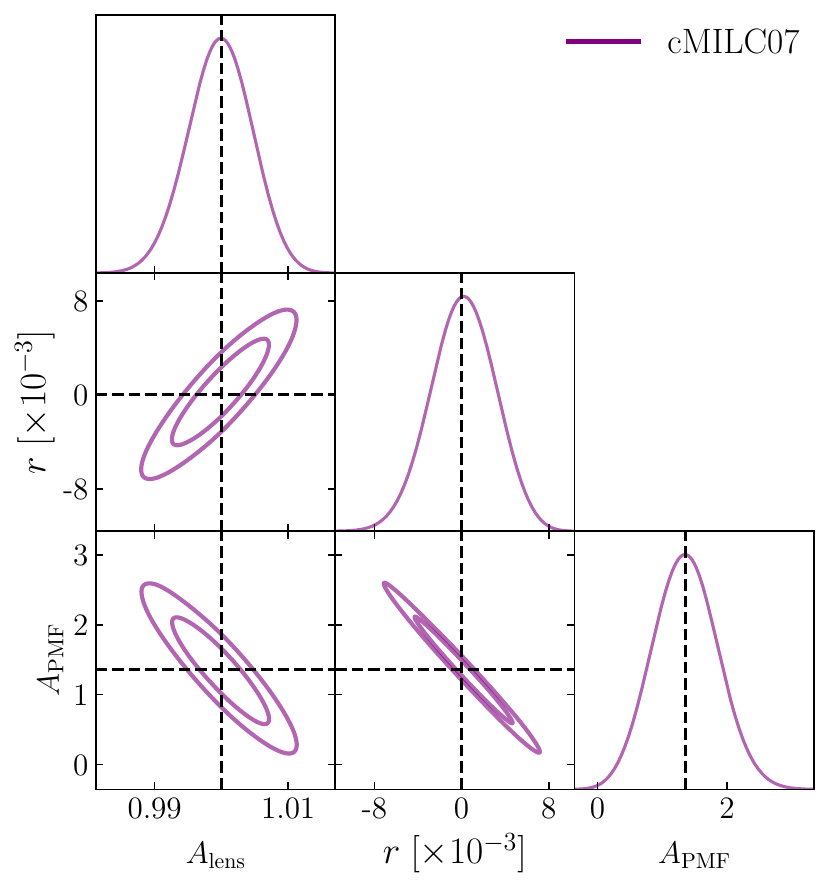}
\caption{Tests of the three-component CMB model $r$+$A_{\rm lens}$+$A_{\rm PMF}$ with non-Gaussian polarized foregrounds (test V in Table \ref{testtype}). The multipole range for this analysis is $20<\ell<1000$. } \label{three_component_comparison}
\end{figure}

The tests in Figs. \ref{bbnull} and \ref{bbsignal} only consider a two-component CMB model, i.e., $r+A_{\rm lens}$, with $A_{\rm lens}$ fixed in the presence of non-Gaussian polarized foreground. Moreover, we consider the parameter inference when $A_{\rm lens}$ is free (test III in Table \ref{testtype}). This result is shown in Fig. \ref{mapcheck2}. We further extend the test to a three-component CMB model with the PMF $B$-mode included (test IV in Table \ref{testtype}), and generate mock data for the three-component CMB model with lensing amplitude fixed at $A_{\rm lens}=1$ but in the presence of non-Gaussian polarized foreground (test IV in Table \ref{testtype}). The two dimensional confidence levels of $r\mbox{-}A_{\rm PMF}$ are inferred after applying the cMILC07 foreground removal procedure. As shown in Fig. \ref{threesignals}, the inferred $r$ and $A_{\rm PMF}$ parameters are consistent with the fiducial values and the cosmological signals are unbiasedly recovered in the presence of non-Gaussian polarized foreground even if there is a strong degeneracy between the PGW and PMF. The result for test V is shown in Fig. \ref{three_component_comparison} where the three free parameters are recovered with the cMILC07 foreground removal.

\begin{figure*}
\includegraphics[width=8cm,height=5.5cm]{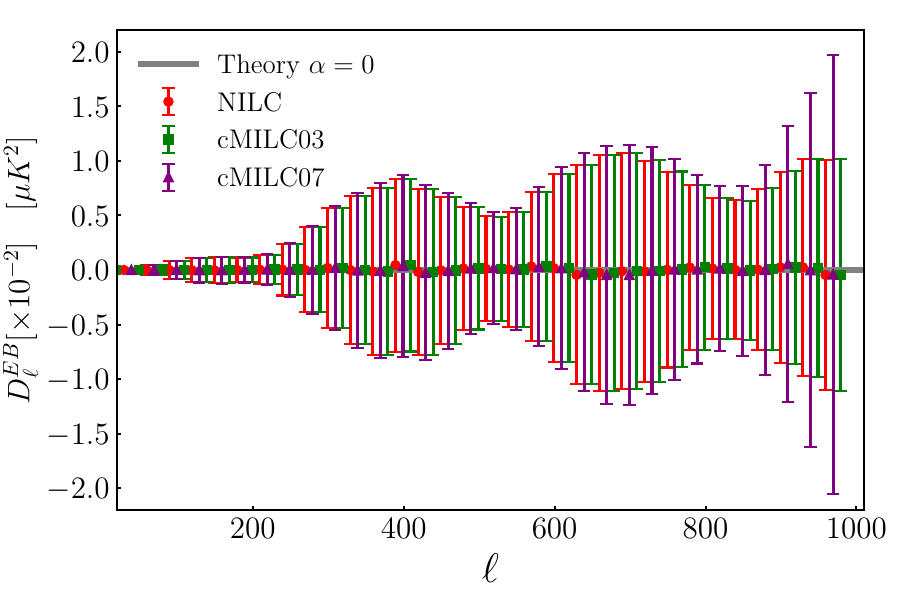}
\includegraphics[width=8cm,height=5.5cm]{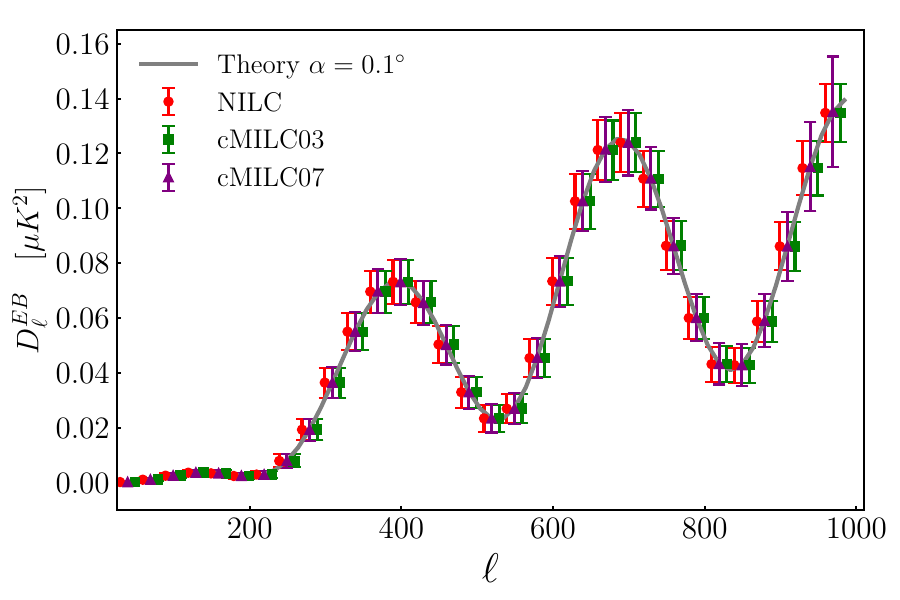}
\caption{Reconstructed $EB$ cross-power spectra from different foreground removal methods when the rotation angles are $\alpha=0$ (left) and $\alpha=0.1^{\circ}$ (right). All the power spectra are consistent with zero. The error bars for NILC and cMILC03 are shifted by $\Delta \ell=10$ for visualization purposes.}\label{ebnull}
\end{figure*}

The tests above validate that the cMILC method can achieve unbiased measurements of the $C_{\ell}^{BB}$ power spectra. However, the non-Gaussian polarized foreground may also affect the cross-power spectrum $C_{\ell}^{EB}$. Thus, we perform similar tests for the $C_{\ell}^{EB}$ power spectrum as done for the $C_{\ell}^{BB}$ power spectrum, i.e., testing both scenarios with no rotation angle $\alpha=0$ and with a rotation angle $\alpha=0.1^{\circ}$. In Fig. \ref{ebnull}, we show the $C_{\ell}^{EB}$ power spectra that are derived from the cross correlations between the reconstructed $E$- and $B$-mode maps using the three foreground removal methods in Fig. \ref{skymapaftercsEB}. All the power spectra are consistent with zero, indicating that the non-Gaussianity does not affect the cross-power spectra. In particular, the biases are zero and band-power error bars are almost the same except that cMILC07's error bars are slightly larger. 
When there is a rotation angle $\alpha=0.1^{\circ}$, the reconstructed cross-power spectra in Fig. \ref{ebnull} are also consistent with the input theory in gray. Similarly, both the biases and errors are slightly affected by the foreground non-Gaussianity. This again confirms that the impact of the foreground non-Gaussianity on the cross-power spectrum can be ignored.

\subsubsection{Comparisons with previous results}
\label{validations}

In order to crosscheck analysis pipeline using the map-space method, we generated mock data with {\picocmb} and {\litebird} specifications~\cite{pico, litebird} to validate the needlet implementations of the ILC series. This mock dataset is referred to as \texttt{PySM simulation} as listed in Table \ref{simtype} and is only limited to the validation discussions in this section. Different from the perturbative expansion of the foreground fluctuations in Eq. (\ref{FGtaylor}), we simulated mock sky observations using the procedures as~\cite{cmilc} and ran \texttt{PySM} to generate the polarized foreground simulations with ``d10" and ``s5" models for dust and synchrotron emissions, respectively. 

We calculated the multifrequency power spectra of the polarized foreground and used them to cross-check the foreground models discussed in Sec. \ref{skysec}. The predicted foreground SEDs in Eqs. (\ref{seddust}) and (\ref{sedsyn}) and the foreground power spectra $C_{\ell}^{{\rm fg},\nu\times\nu',0\times 0}$ (Eq. (\ref{fghigherALL})) with the fiducial parameters are consistent with the \texttt{PySM} mock data.

We synthesized the mock data for {\picocmb} and {\litebird} with the \texttt{PySM}-generated foreground, white noise and Gaussian CMB realizations that are generated from lensed CMB power spectra by \texttt{CAMB}~\cite{camb}. The CMB $B$-mode signals for this validation only contain lensing contributions with no inflationary $B$-mode signal ($r=0$). The white noise maps with {\picocmb} and {\litebird} noise levels were added in the mock data and frequency-dependent FWHMs were also incorporated. In this work, the noise maps were deconvolved with different beam profiles before the foreground removal. 

The {\picocmb} and {\litebird} sky maps for the polarization $Q$ and $U$ were converted to $E$ and $B$ modes with spin-2 spherical harmonic transformations. Then the $E$- and $B$-modes were converted to spin-0 maps while adjusting to the same resolution at 40 {\rm arcmin} for all frequencies. Then we applied different foreground removal methods including NILC, cMILC03, cMILC06 and cMILC08 to the mock data and used \texttt{NaMaster}~\cite{namaster} to calculate band powers of the component-separated polarization maps for the $\ell$-range $20<\ell<1000$ with a binning size $\Delta \ell=16$. We obtained the foreground residuals and noise estimates which agree with the results in~\cite{cmilc}.

\section{Conclusions}

The CMB $B$-mode is an important probe to fundamental physics. Especially, the $B$-mode signal from primordial gravitational waves (PGWs) can provide crucial information for the high-energy scale in the early Universe. However, the $B$-mode signal is a multi-component signal which can originate from gravitational lensing effects and Galactic foreground contaminants. These two sources can contribute dominant $B$-mode signals from small to large angular scales, making the PGW detection challenging.

In addition to these known sources which have been detected from CMB observations, we investigated two different $B$-mode signals in this work. 
The spatially varying components of the foreground SEDs can create excess foreground $B$-mode power at large angular scales. Also, the tensor perturbations of the PMF can create a $B$-mode signal degenerate with the PGW. It is thus of great importance to investigate the inference of the PGW $B$-mode signal in the presence of both contamination.

The degeneracy between PGW and PMF $B$-mode signals has been studied in~\cite{degeneracy2, degeneracy2} and can be broken by extending the $B$-mode power spectrum to a higher $\ell_{\rm max}$ or including a cosmic rotation signals powered by the PMF~\cite{pmf_aa}.  
On the other hand, the foreground SEDs may not be isotropic. If the foreground parameters such as the spectral indices and dust temperature are spatially varying, nonnegligible foreground residuals may still exist even if the standard component-separation methods are applied, because there are higher-order correlations generated by these spatial components. If unaccounted for, both the spatial components of foreground SEDs and the PMF tensor modes might cause nonnegligible biases or additional errors to the tensor-to-scalar ratio.

In this work, we investigated the impact of both sources on the PGW inference using both the map- and power-spectrum-based methods. The map-space methods considered in this work include the standard ILC and the recent cMILC methods, which can achieve the minimum variance for the component-separated maps. In addition, the ILC series were implemented with the needlet configurations. 
We also built power-spectrum models for CMB, instrumental noise and polarized foreground at different perturbation levels. The CMB $B$-mode power spectrum consists of the PGW, lensing and PMF components. A Bayesian analysis framework was established for multifrequency mock power spectra among different frequency bands. 

For the CMB sky with no PGW, we performed the Bayesian analysis to infer all the foreground parameters and the tensor-to-scalar ratio using the power-spectrum-based method. The posterior distribution function (PDF) of $r$ from the power-spectrum-based method is consistent with that of the map-based method. We applied both methods to the mock data with PMF signal included and assumed a CMB model with free parameters $r, A_{\rm lens}, A_{\rm PMF}$. For the power-spectrum-based method, we found that a larger $\ell_{\rm max}$ can effectively break the degeneracy between $r$ and $A_{\rm PMF}$ while the two parameters are weakly correlated with the foreground parameters. The two dimensional contours of the cosmological parameters inferred from either the reconstructed maps or the multifrequency power spectra are found to be consistent with each other.

For this work, we created the end-to-end simulations at $20\le \nu\le 350$ GHz with seven frequency bands for future CMB polarization experiments following the sky model described in Eq (\ref{sky}) and studied how the spatial SED moments and the PMF can affect the PGW inference.

Using the map-based methods, we reconstructed the polarization maps from the seven frequency bands assuming a CMB model with and without PGW signals. All the power spectra of the foreground residuals and noise biases were calculated and we found that the foreground residuals can be gradually suppressed from NILC, cMILC03 to cMILC07, corresponding to more foreground moments, as seen from Figs. \ref{bbnull} and \ref{bbsignal}. But this will incur the noise penalty and the noise biases of the ILC variants with more moments are much higher. For future CMB polarization experiments, similar tests can be performed to determine optimal moment combinations for a specific scenario. 

In addition, we made CMB-only simulations with a three-component model, i.e., $r+A_{\rm lens}+A_{\rm PMF}$ with $A_{\rm lens}$ fixed. Although there is a strong degeneracy between the tensor-to-scalar ratio $r$ and PMF amplitude $A_{\rm PMF}$ as found in~\cite{degeneracy, degeneracy2}, the inferred parameters $r$ and $A_{\rm PMF}$ are consistent with the fiducial values. Moreover, if the non-Gaussian polarized foreground is added into the simulations, the reconstructed maps with cMILC07 can still yield unbiased parameter inference in Fig. \ref{threesignals}. 

In Fig. \ref{ebnull}, we investigated the $EB$ power-spectrum calculations with two different rotation angles, which can be precisely extracted with no significant biases. Thus, the cross-power spectrum is quite immune to the foreground non-Gaussianity. We validated the implementations of the map-space methods using {\picocmb} and {\litebird} mock data~\cite{pico, litebird}. The resulting foreground residuals and noise biases are consistent with the results in~\cite{cmilc}.

The future CMB polarization experiment will be integrated with both low- and high-frequency channels to better control the Galactic foreground. The non-Gaussian features in the polarized foreground will be a new challenge to the study of the PGW and other exotic $B$-mode signatures. On the other hand, other cosmological origins of the $B$-mode signal, such as the tensor mode of the PMF, may further complicate the PGW detection. This work provides clear guidance for the next-generation CMB polarization experiments which could achieve unprecedented high-sensitivity to distinguish different cosmological $B$-mode signals from other complex astrophysical signals.

\acknowledgments

We are grateful for the helpful discussions with Jacques Delabrouille, Mathieu Remazeilles, Pengjie Zhang and Tom Crawford. This work is supported by the starting grant of USTC. F.B.A. acknowledges the support of the CAS. We acknowledge the use of the \healpix~\cite{2005ApJ...622..759G} and \namaster~\cite{namaster} packages.

\bibliography{randpmf}
\end{document}